\documentclass{emulateapj} 

\gdef\1054{MS\,1054--03}
\gdef\Lya{Ly$\alpha$}
\gdef\Ha{H$\alpha$}
\def\farcs{\hbox{$.\!\!^{\prime\prime}$}}
\def\simgeq{{\raise.0ex\hbox{$\mathchar"013E$}\mkern-14mu\lower1.2ex\hbox{$\mathchar"0218$}}} 

\def\placefigureOne{
\begin{figure}[htbp]
\centering
\plotone{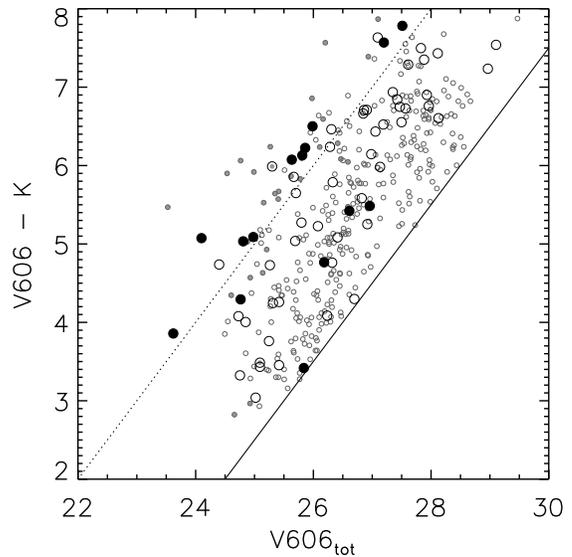} 
\caption{\small
Sample selection for the spectroscopic survey of DRGs.
The location of all DRGs with $K_{s,tot}<22.5$ in the HDFS, \1054, and
CDFS fields is plotted with small circles in the $V_{606}-K_s$ versus
$V_{606,tot}$ color-magnitude diagram.  Large circles represent DRGs
observed during the spectroscopic campaign described in this paper,
with filled black symbols indicating the successful redshift
determinations.  Filled grey circles are DRGs in the CDFS for which a
spectroscopic redshift is available from the literature.  Lines of
constant $K_{s,tot}=22.5$ (the magnitude limit of our sample; {\it
solid}) and $K_{s,tot}=20$ ({\it dotted}) are plotted to guide the
eye.  The sample targeted by our survey shows a representative range
in $V_{606}-K_s$ and in $V_{606,tot}$.  The success rate is biased
toward DRGs that are bright in the $K_s$-band.
\label {VKvsV.fig}
}
\end{figure}
}

\def\placefigureTwo{
\begin{figure*}[htbp]
\centering
\plotone{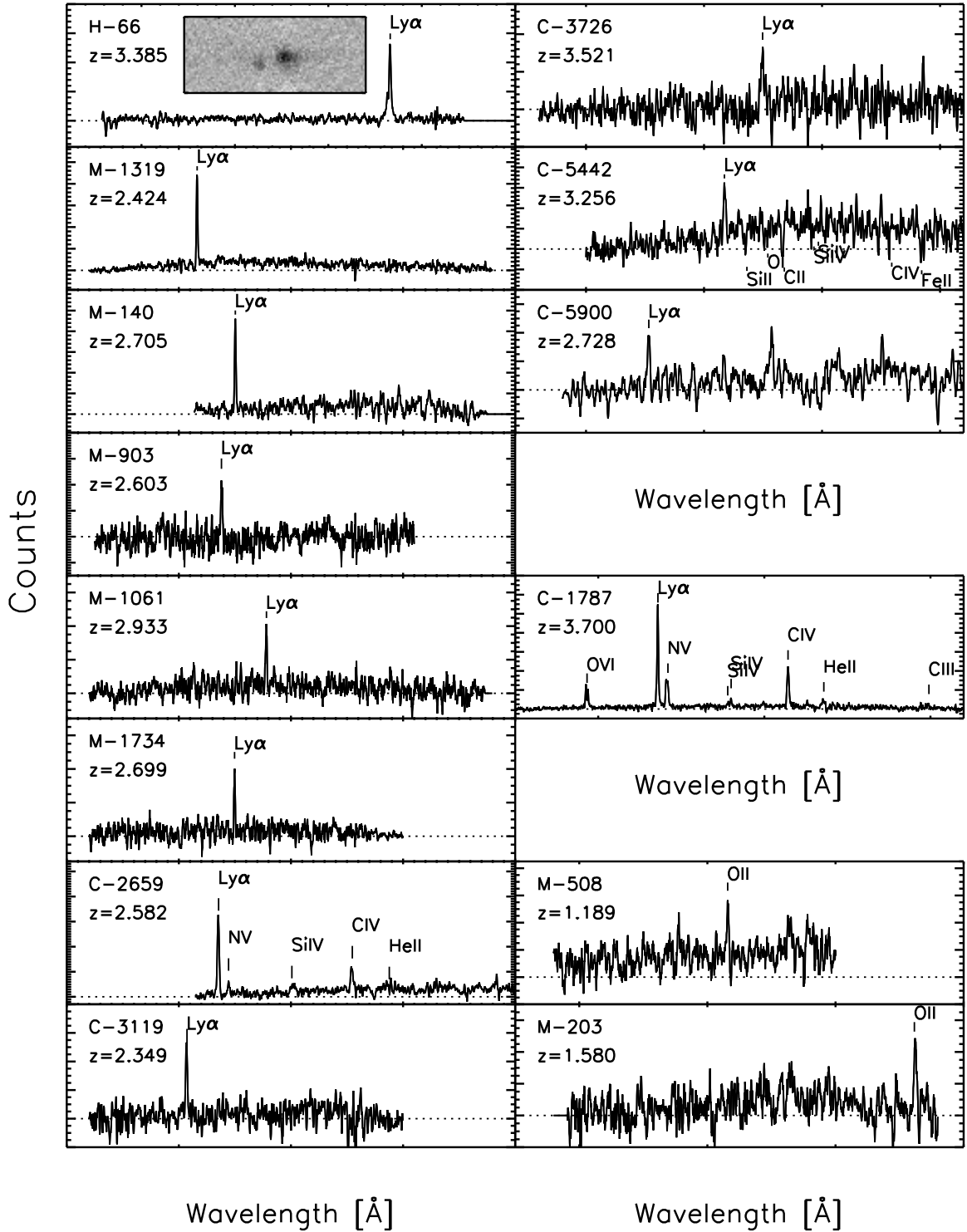} 
\caption{\small
1D optical spectra of DRGs observed in our survey with
successful redshift identification.  The presented spectra of DRGs at
$z>2$ show \Lya\ in emission, possibly in combination with other lines.
Two interlopers at $z<2$ were identified by the presence of [OII]3727
in emission, with the continuum extending blueward of the emission
line.  Inset for object H-66 is a part of the GMOS 2D spectrum,
showing a smaller feature close to the \Lya\ emission from the target.
Galaxies C-1787 and C-2659 show evidence of AGN activity in their
optical spectra.  Interstellar absorption lines are detected in
C-5442.
\label{oneDspec.fig}
}
\end{figure*}
}

\def\placefigureThree{
\begin{figure}[htbp]
\centering
\plotone{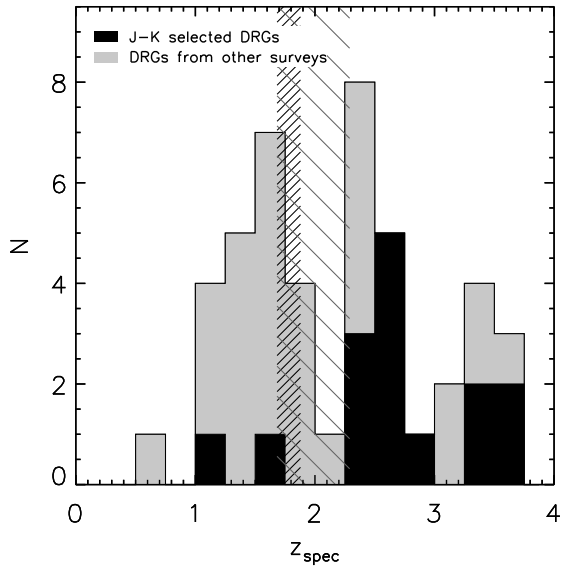} 
\caption{\small Spectroscopic redshift histogram of DRGs in the HDFS,
\1054, and the CDFS.  Redshifts obtained for purely $J-K > 2.3$
selected galaxies are presented in black.  Additional spectroscopic
redshifts of objects satisfying $J-K>2.3$ from other surveys are
indicated in dark grey.  Their redshift distribution is different,
owing to the different criteria used to select them.  The closely and
widely hatched regions mark the range in redshifts where both
[OII]3727 and \Lya\ fall outside the sensitive part of the LRIS and
FORS2 detectors respectively.
\label {speczhist.fig}
}
\vspace{-4mm}
\end{figure}
}

\def\placefigureFour{
\begin{figure}[t]
\centering
\plotone{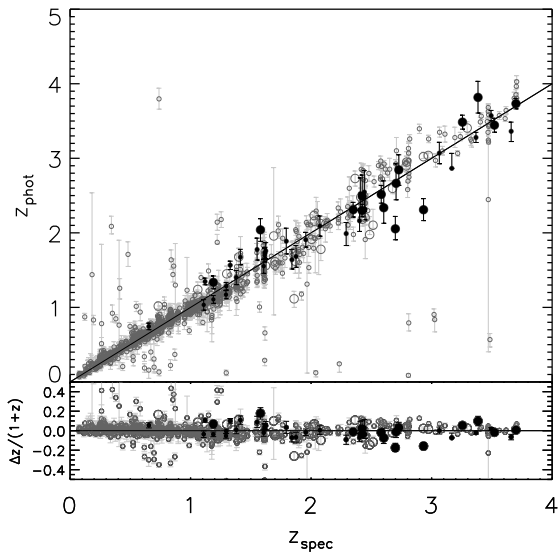} 
\caption{\small
Direct comparison between photometric and
spectroscopic redshifts for all sources with $K_{s,tot} < 22.5$ in the
HDFS, \1054, and CDFS fields for which a reliable spectroscopic
redshift is available.  Distant Red Galaxies are highlighted in black.
Large symbols denote redshifts obtained during our spectroscopic
survey.
\label {photz_specz.fig}
}
\vspace{-4mm}
\end{figure}
}

\def\placefigureFive{
\begin{figure}[b]
\centering
\plotone{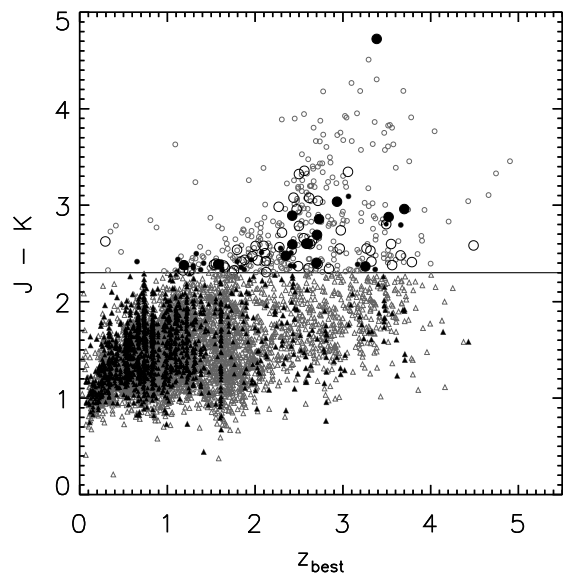} 
\caption{\small
$J-K$ versus redshift for all sources with $K_{s,tot}
< 22.5$ in the HDFS, \1054, and CDFS fields.  Filled symbols
are used for spectroscopic redshifts.  For other sources the
photometric redshift estimate.  Large symbols represent galaxies selected for our
spectroscopic survey.  Objects above the horizontal line marking
$J-K=2.3$ satisfy the DRG criterion.  Selecting galaxies based on
their red $J-K$ color is an efficient means to find $z>2$ galaxies.
\label {JKvsz.fig}
}
\end{figure}
}

\def\placefigureSix{
\begin{figure}[t]
\centering
\plotone{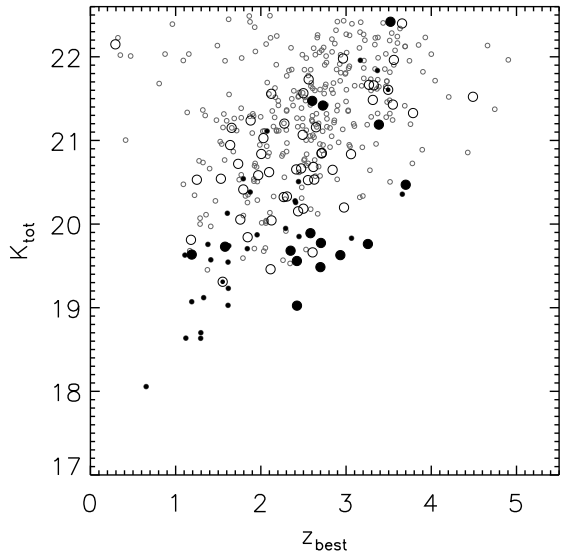} 
\caption{\small Observed $K_s$-band magnitude versus redshift for all
DRGs with $K_{s,tot} < 22.5$ in the HDFS, \1054, and CDFS fields.
Filled circles are used for DRGs with spectroscopic redshifts.  For
other DRGs ({\it empty circles}) the photometric redshift estimate is
plotted.  Large symbols represent galaxies in our spectroscopic
survey.  Low-redshift DRGs reach to brighter $K_{s,tot}$ than
high-redshift DRGs.
\label {Kvsz.fig}
}
\vspace{-4mm}
\end{figure}
}

\def\placefigureSeven{
\begin{figure}[t]
\centering
\plotone{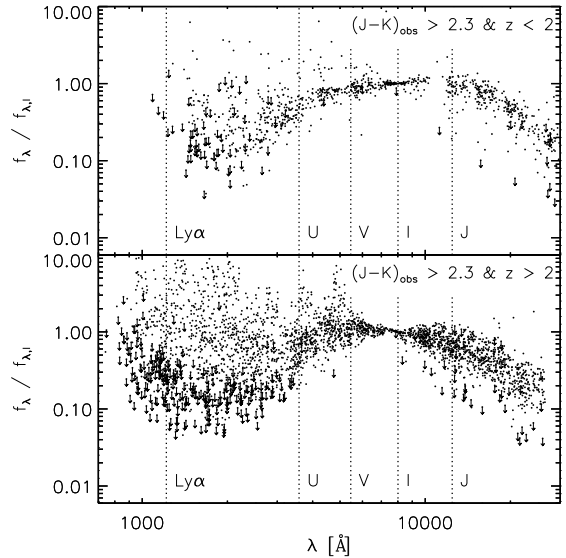} 
\caption{\small
{\it Top panel:} Rest-frame broad-band SEDs,
normalized to the rest-frame $I$-band flux, of all low-redshift
($z<2$) DRGs to $K_{s,tot}<22.5$ in the HDFS, \1054, and CDFS fields.
{\it Bottom panel:} High-redshift ($z>2$) DRGs to the same magnitude
limit.  Upper limits indicate the 1$\sigma$ confidence levels.
Low-redshift DRGs have a red SED shape from the rest-frame UV to the
rest-frame $J$-band, whereas the SEDs of high-redshift DRGs show a
wide range in rest-frame UV slopes and are on average declining
redward of the rest-frame $V$-band.
\label {restSED.fig}
}
\vspace{-4mm}
\end{figure}
}

\def\placefigureEight{
\begin{figure}[t]
\centering
\plotone{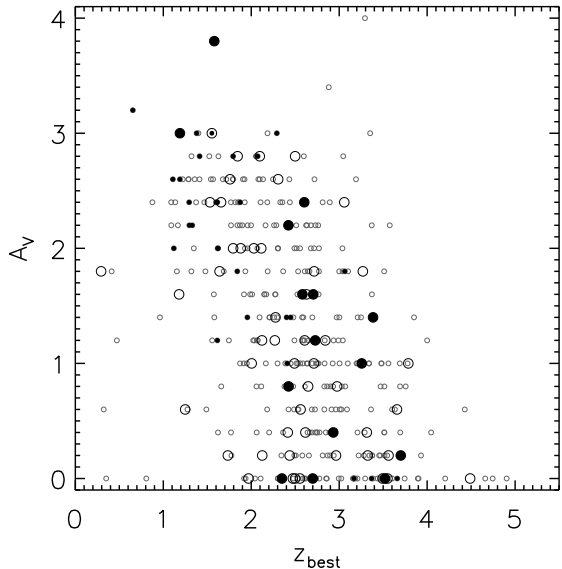} 
\caption{\small
Best-fitted $A_V$ versus redshift ($z_{phot}$ or
$z_{spec}$ when available) for all DRGs with $K_{s,tot} < 22.5$ in the
HDFS, \1054, and CDFS fields.  Spectroscopic redshifts are marked with
filled symbols.  Large symbols indicate galaxies that were part of our
spectroscopic follow-up of DRGs.  The dust content of DRGs decreases
with increasing redshift.
\label {Avvsz.fig}
}
\vspace{-2mm}
\end{figure}
}

\def\placetableOne{
\begin{deluxetable*}{lcccc}
\tablecolumns{5}
\tablewidth{0pc}
\tablecaption{Spectroscopic redshifts for DRGs from cross-correlation with other surveys in the CDFS \label{others.tab}
}
\tablehead{
\colhead{ID} & \colhead{ra} & \colhead{dec} & \colhead{$z_{spec}$} & \colhead{Source}
}
\startdata
C-1187 & 53.0603067 & -27.8760705 & 1.875 & Fadda et al. (in prep) \\
C-1191 & 53.1824218 & -27.8753155 & 1.624 & Fadda et al. (in prep) \\
C-1553 & 53.0784636 & -27.8598817 & 3.660 & CXO \\
C-1957 & 53.1988252 & -27.8438850 & 1.615 & Kriek et al.\tablenotemark{a} \\
C-1978 & 53.0716662 & -27.8436585 & 1.956 & Fadda et al. (in prep) \\
C-2239 & 53.0726960 & -27.8342035 & 1.618 & Fadda et al. (in prep) \\
C-2482 & 53.2021505 & -27.8263119 & 1.120 & VLT/FORS2 \\
C-2792 & 53.0667164 & -27.8165291 & 1.413 & VLT/FORS2 \\
C-2855 & 53.1652224 & -27.8140093 & 3.064 & CXO \\
C-3129 & 53.0446457 & -27.8019901 & 0.654 & K20 \\
C-3521 & 53.0759416 & -27.7858175 & 1.379 & Roche et al. 2006 \\
C-3854 & 53.0940038 & -27.7608387 & 2.073 & Fadda et al. (in prep) \\
C-3886 & 53.0228760 & -27.7570050 & 1.189 & VLT/FORS2 \\
C-3968 & 53.1729054 & -27.7444701 & 1.296 & VLT/FORS2 \\
C-3973 & 53.0936462 & -27.7440207 & 3.494 & VLT/VIMOS \\
C-4679 & 53.1133547 & -27.6977781 & 3.168 & VLT/VIMOS \\
C-4712 & 53.0632815 & -27.6996566 & 2.402 & CXO \\
C-4735 & 53.1375292 & -27.7001314 & 2.448 & GMASS (Kurk et al. in prep) \\
C-4982 & 53.1496306 & -27.7113616 & 1.615 & GMASS (Cappellari et al. 2009) \\
C-5097 & 53.0508272 & -27.7137057 & 2.442 & Fadda et al. (in prep) \\
C-5162 & 53.0523044 & -27.7183135 & 1.794 & Fadda et al. (in prep) \\
C-5177 & 53.1070458 & -27.7181950 & 2.291 & CXO \\ 
C-5257 & 53.0793872 & -27.7208650 & 2.408 & GMASS (Kurk et al. in prep) \\
C-5605 & 53.1205657 & -27.7365600 & 3.368 & MUSYC IMACS\tablenotemark{b} \\
C-5724 & 53.0963554 & -27.7450675 & 1.607 & GMASS (Kurk et al. 2009) \\
C-5842 & 53.0362490 & -27.7522039 & 1.294 & K20 \\
C-6063 & 53.0732969 & -27.7643758 & 1.841 & Fadda et al. (in prep) \\
C-6070 & 53.0266661 & -27.7652460 & 1.329 & VLT/FORS2 \\
C-6132 & 53.1169241 & -27.7684461 & 1.109 & K20 \\
C-6161 & 53.1655522 & -27.7698397 & 1.552 & GMASS (Kurk et al. in prep)
\enddata
\tablenotetext{a}{SINFONI spectroscopy, December 2006, program 076.A-0464.}
\tablenotetext{b}{Optical IMACS spectroscopy by the MUSYC survey from private communication.}
\end{deluxetable*}
}

\def\placetableTwo{
\vspace{-1in}
\begin{deluxetable}{lcccc}
\tablecolumns{5}
\tablewidth{0pc}
\tablecaption{Spectroscopic redshifts for non-DRGs obtained during our spectroscopic survey \label{nonDRG.tab}
}
\tablehead{
\colhead{ID\tablenotemark{a}} & \colhead{ra} & \colhead{dec} & \colhead{$z_{spec}$} & \colhead{Remark\tablenotemark{b}}
}
\startdata
H-92  &  338.22568 &  -60.569154 & 2.412    & $I_{814}-H = 1.84$; $J_s-K_s = 1.54$ \\
H-228 &  338.21679 &  -60.561796 & 3.295    & $I_{814}-H = 1.35$; $J_s-K_s = 1.69$ \\
H-245 &  338.22862 &  -60.561701 & 2.676    & $I_{814}-H = 0.97$; $J_s-K_s = 1.24$ \\
H-257 &  338.21121 &  -60.557914 & 2.027    & $I_{814}-H = 2.19$; $J_s-K_s = 1.49$ \\
H-290 &  338.26335 &  -60.558267 & 2.025    & $I_{814}-H = 2.11$; $J_s-K_s = 1.37$ \\
H-294 &  338.27042 &  -60.558536 & 2.365    & $I_{814}-H = 1.64$; $J_s-K_s = 1.78$ \\
H-408 &  338.24993 &  -60.551115 & 1.228    & $I_{814}-H = 1.80$; $J_s-K_s = 1.29$ \\
H-470 &  338.22038 &  -60.554717 & 1.284    & $I_{814}-H = 2.94$; $J_s-K_s = 2.01$ \\
H-565 &  338.22220 &  -60.544237 & 1.114    & $I_{814}-H = 2.39$; $J_s-K_s = 1.81$ \\
H-620 &  338.23714 &  -60.536690 & 1.558    & $I_{814}-H = 1.58$; $J_s-K_s = 1.26$ \\
H-657 &  338.20360 &  -60.531616 & 2.793    & $I_{814}-H = 2.09$; $J_s-K_s = 1.91$ \\
H-806 &  338.20579 &  -60.540609 & 2.789    & $I_{814}-H = 1.20$; $J_s-K_s = 1.15$ \\
H- &  338.25705 &  -60.590965 & 0.695    & - \\ 
H- &  338.27145 &  -60.577366 & 0.439    & - \\ 
H- &  338.27145 &  -60.579903 & 0.844    & - \\ 
H- &  338.28201 &  -60.587112 & 0.344    & - \\ 
H- &  338.25686 &  -60.59766  & 2.899    & LBG candidate \\ 
H- &  338.28486 &  -60.57794  & 3.190    & LBG candidate \\ 
M-147 &  164.23573 &  -3.6498842 & 1.265    & $I_{814}-H = 2.45$;  $J_s-K_s = 1.55$ \\
M-161 &  164.24502 &  -3.6475178 & 1.859    & $I_{814}-H = 2.21$;  $J_s-K_s = 1.87$ \\
M-266 &  164.22595 &  -3.6422003 & 2.005    & $I_{814}-H = 1.57$;  $J_s-K_s = 1.08$ \\
M-303 &  164.21742 &  -3.6400908 & 2.486    & $I_{814}-H = 2.03$;  $J_s-K_s = 1.25$ \\
M-383 &  164.22318 &  -3.6365197 & 2.123    & $I_{814}-H = 2.30$;  $J_s-K_s = 1.62$ \\
M-450 &  164.20416 &  -3.6339978 & 0.346    & no $I_{814}$ coverage;  $J_s-K_s = 1.85$ \\
M-713 &  164.24837 &  -3.6252800 & 1.700    & $I_{814}-H = 3.67$;  $J_s-K_s = 1.75$ \\
M-897 &  164.24914 &  -3.6203344 & 2.973    & $I_{814}-H = 1.13$;  $J_s-K_s = 1.31$ \\
M-972 &  164.21320 &  -3.6176475 & 2.448    & $I_{814}-H = 2.01$;  $J_s-K_s = 1.82$ \\
M-1132 &  164.27260 &  -3.6095794 & 1.060    & $I_{814}-H = 3.23$;  $J_s-K_s = 2.14$ \\
M-1155 &  164.22757 &  -3.6094061 & 1.622    & $I_{814}-H = 3.59$;  $J_s-K_s = 1.90$ \\
M-1272 &  164.27786 &  -3.6050289 & 0.829    & $I_{814}-H = 1.31$;  $J_s-K_s = 1.12$ \\
M-1396 &  164.24016 &  -3.6010686 & 2.514    & $I_{814}-H = 2.13$;  $J_s-K_s = 1.65$ \\
M-1450 &  164.24319 &  -3.5979289 & 0.622    & $I_{814}-H = 1.23$;  $J_s-K_s = 1.08$ \\
M-1459 &  164.25297 &  -3.5974653 & 2.081    & $I_{814}-H = 3.92$;  $J_s-K_s = 2.22$ \\
M-1637 &  164.23843 &  -3.5876183 & 1.300    & $I_{814}-H = 3.10$;  $J_s-K_s = 2.24$ \\
M-1728 &  164.26288 &  -3.5815978 & 2.93200    & $I_{814}-H = 1.63$;  $J_s-K_s = 1.42$ \\
M-  &  164.23486 & -3.5825150 & 2.428  &  NB4190 \\ 
M-  &  164.21390 & -3.5891633 & 2.436  &  NB4190 \\ 
M-  &  164.19865 & -3.6408465 & 2.428  &  NB4190 \\ 
M-  &  164.22060 & -3.6178541 & 2.422  &  NB4190 \\ 
M-  &  164.23906 & -3.5812418 & 2.280  &  NB4190 \\ 
M-  &  164.27251 & -3.5855079 & 0.559  &  NB4190 \\ 
M-  &  164.21590 & -3.6068938 & 0.119  &  NB4190 \\ 
M-  & 164.22655  & -3.6836915 & 0.261  &  - \\ 
M-  & 164.22023  & -3.6792324 & 1.086  &  - \\ 
M-  & 164.22426  & -3.6761484 & 0.577  &  - \\ 
C-2363 &  53.082743 &  -27.831706 & 0.246    & $I_{775}-H = 1.91$;  $J-K_s = 1.15$ \\
C-2472 &  53.093660 &  -27.826402 & 0.732    & $I_{775}-H = 3.11$;  $J-K_s = 2.18$ \\
C-2484 &  53.092048 &  -27.827811 & 0.731    & $I_{775}-H = 1.33$;  $J-K_s = 0.96$ \\
C-3358 &  53.178065 &  -27.792739 & 1.427    & $I_{775}-H = 3.54$;  $J-K_s = 2.17$ \\
\enddata 
\tablenotetext{a}{H- stands for HDFS, M- for \1054, and C- for CDFS.  Objects without ID number are either located outside the area covered by the $K_s$-selected catalog or are not detected in $K_s$.}
\tablenotetext{b}{Objects with a narrow-band flux excess at 4190 \AA\ are indicated with NB4190.}
\end{deluxetable}
}

\def\placetableThree{
\begin{deluxetable*}{lllllll} 
\tablecolumns{5}
\tablewidth{0pc}
\tablecaption{Spectroscopic observing runs \label{observations.tab}
}
\tablehead{
\colhead{Date} & \colhead{Telescope} & \colhead{Instrument} & \colhead{Field} & \colhead{Total exposure time} & \colhead{Instrument settings} & \colhead{Seeing} \\
\colhead{}     & \colhead{}          & \colhead{}           & \colhead{}      & \colhead{s}                   & \colhead{}                    & \colhead{``}
}
\startdata
February 2002  &  Keck  &  LRIS  &  \1054  &  72000  &  D680 dichroic  &  0.8 - 1.5\tablenotemark{p} \\
               &        &        &         &         &  blue: 300 line mm$^{-1}$ &  \\
               &        &        &         &         &  red: 400/8500 \AA\ and 600/1 $\mu$m grating &  \\
September 2002 &  VLT   &  FORS2 &  HDFS   &  19800  &  GRIS\_300V, filter gg375  & 0.8 - 2.0\tablenotemark{p} \\
December 2002  &  VLT   &  FORS2 &  CDFS   &  29700  &  GRIS\_300V                & 1.0 - 2.3\tablenotemark{p} \\
January 2003   &  Keck  &  LRIS  &  \1054  &  6800   &  D680 dichroic            & 0.7 - 0.8 \\
               &        &        &         &         &  blue: 400/3400 \AA\ grism  & \\
               &        &        &         &         &  red: 400/8500 \AA\ grating & \\
               &        & DEIMOS &  \1054  &  18000  &  mask1: 600/7300 \AA\ grism, filter gg495  & 0.8 - 1.0 \\
               &        &        &         &  36240  &  mask2,3: 600/7700 \AA\ grism, filter og550  & 0.7 - 1.4 \\
March 2003     &  Keck  & LRIS   &  \1054  &  14400  &  D560 dichroic            & 0.9 - 1.1\tablenotemark{p} \\
               &        &        &         &         &  blue: 400/3400 \AA\ grism  & \\
               &        &        &         &         &  red: 400/8500 \AA\ grating & \\
March 2003     &  VLT   & FORS2  &  \1054  &  14400  &  GRIS\_300V, filter gg375  & 0.6 - 0.9\tablenotemark{p} \\
September 2003 &  Gemini-South & GMOS & HDFS & 38400 &  B600/4500 \AA\ and B600/4530 \AA\ grating & 0.9 - 1.4 \\
October 2003   &  VLT   & FORS2  &  CDFS   &  24470  &  GRIS\_300V                & 0.5 - 2.0\tablenotemark{p} \\
               &        &        &  HDFS   &  16200  &  GRIS\_300V                & 0.65 - 1.8\tablenotemark{p} \\
November 2003  &  Keck  & LRIS   &  CDFS   &   9300  &  D560 dichroic            & 0.7 - 1.5\tablenotemark{p} \\
               &        &        &         &         &  blue: 400/3400 \AA\ grism  & \\
               &        &        &         &         &  red: 400/8500 \AA\ grating & 
\enddata
\tablenotetext{p}{Observing conditions were photometric.}
\end{deluxetable*}
}

\def\placetableFour{
\begin{deluxetable*}{lcccc}
\tablecolumns{5}
\tablewidth{0pc}
\tablecaption{Spectroscopic redshifts from our spectroscopic follow-up of DRGs \label{ours.tab}
}
\tablehead{
\colhead{ID\tablenotemark{a}} & \colhead{ra} & \colhead{dec} & \colhead{$z_{spec}$} & \colhead{Remark}
}
\startdata
H-66 & 338.2713649 & -60.5703250 & 3.385 & has close companion at 2.6 kpc\\
M-140 & 164.2106125 & -3.6508417 & 2.705 & -\\
M-203 & 164.2078833 & -3.6463678 & 1.580 & -\\
M-508 & 164.2299500 & -3.6315592 & 1.189 & -\\
M-903 & 164.1998917 & -3.6207567 & 2.603 & -\\
M-1061 & 164.2394875 & -3.6131875 & 2.933 & optical and NIR flux offset by $1\farcs 5$\\
M-1319 & 164.2775375 & -3.6010592 & 2.424 & -\\
M-1383 & 164.2603167 & -3.6006669 & 2.423 & redshift from NIR spectroscopy\\
M-1734 & 164.2233917 & -3.5811008 & 2.699 & -\\
C-1787 & 53.1243363 & -27.8516408 & 3.700 & also analysed by Norman et al. (2002)\\
C-2659 & 53.1488159 & -27.8211517 & 2.582 & -\\
C-3119 & 53.1231066 & -27.8033550 & 2.349 & -\\
C-3726 & 53.0550864 & -27.7785031 & 3.521 & -\\
C-5442 & 53.1177728 & -27.7342424 & 3.256 & -\\
C-5900 & 53.1080817 & -27.7539822 & 2.728 & -
\enddata
\tablenotetext{a}{H- stands for HDFS, M- for \1054, and C- for CDFS.}
\end{deluxetable*}
}

\def\placetableFive{
\begin{deluxetable*}{lccccccc}
\tablecolumns{8}
\tablewidth{0pc}
\tablecaption{Quality of photometric redshifts: statistical measures of $\Delta z/(1+z)$ \label{photozqual.tab}
}
\tablehead{
\colhead{Sample} & \colhead{Median} & \colhead{$\sigma_{NMAD}$} & \colhead{Percentage of catastrophic ($> 5 \sigma$) outliers}
}
\startdata
DRGs (purely $J-K$ selected)               & -0.011  & 0.054 & 0.0\\
DRGs (purely $J-K$ selected, $z_{spec}>2$) & -0.012 & 0.047 & 0.0\\
DRGs (all)                                 & 0.001  & 0.056 & 0.0\\
DRGs (all, $z_{spec}>2$)                   & -0.014 & 0.048 & 0.0\\
All galaxies                               & 0.001  & 0.034 & 3.0\\  
All galaxies $z_{spec}>2$                  & 0.005  & 0.055 & 4.8
\enddata
\end{deluxetable*}
}

\begin {document}

\title {Optical spectroscopy of Distant Red Galaxies}

\author{Stijn Wuyts\altaffilmark{1,2}, Pieter G. van Dokkum\altaffilmark{3}, Marijn Franx\altaffilmark{4}, Natascha M. F\"{o}rster Schreiber\altaffilmark{5}, Garth D. Illingworth\altaffilmark{6}, Ivo Labb\'{e}\altaffilmark{7}, Gregory Rudnick\altaffilmark{8}}
\altaffiltext{1}{Harvard-Smithsonian Center for Astrophysics, 60 Garden Street, Cambridge, MA 02138}
\altaffiltext{2}{W. M. Keck Postdoctoral Fellow}
\altaffiltext{3}{Department of Astronomy, Yale University, New Haven, CT 06520-8101}
\altaffiltext{4}{Leiden Observatory, Leiden University, P.O. Box 9513, NL-2300 RA, Leiden, The Netherlands}
\altaffiltext{5}{MPE, Giessenbackstrasse, D-85748, Garching, Germany}
\altaffiltext{6}{UCO/Lick Observatory, University of California, Santa Cruz, CA 95064}
\altaffiltext{7}{Hubble Fellow, Carnegie Observatories, 813 Santa Barbara Street, Pasadena, CA 91101}
\altaffiltext{8}{Department of Physics and Astronomy, University of Kansas, Lawrence, KS 66045}

\begin{abstract}
We present optical spectroscopic follow-up of a sample of Distant Red
Galaxies (DRGs) with $K^{tot}_{s,Vega}<22.5$, selected by
$(J-K)_{Vega}>2.3$, in the Hubble Deep Field South (HDFS), the \1054
field, and the Chandra Deep Field South (CDFS).  Spectroscopic
redshifts were obtained for 15 DRGs.  Only 2 out of 15 DRGs are
located at $z<2$, suggesting a high efficiency to select high-redshift
sources.  From other spectroscopic surveys in the CDFS targeting
intermediate to high redshift populations selected with different
criteria, we find spectroscopic redshifts for a further 30 DRGs.  We
use the sample of spectroscopically confirmed DRGs to establish the
high quality (scatter in $\Delta z/(1+z)$ of $\sim 0.05$) of their
photometric redshifts in the considered deep fields, as derived with
EAZY (Brammer et al. 2008).  Combining the spectroscopic and
photometric redshifts, we find that 74\% of DRGs with
$K^{tot}_{s,Vega} < 22.5$ lie at $z>2$.  The combined spectroscopic
and photometric sample is used to analyze the distinct intrinsic and
observed properties of DRGs at $z<2$ and $z>2$.  In our photometric
sample to $K^{tot}_{s,Vega} < 22.5$, low-redshift DRGs are brighter in
$K_s$ than high-redshift DRGs by 0.7 mag, and more extincted by 1.2
mag in $A_V$.  Our analysis shows that the DRG criterion selects
galaxies with different properties at different redshifts.  Such
biases can be largely avoided by selecting galaxies based on their
rest-frame properties, which requires very good multi-band photometry
and high quality photometric redshifts.
\end{abstract}


\section {Introduction}
\label{intro.sec}
Studies of the history of star formation and
mass assembly in galaxies requires samples of galaxies over a range of
lookback times.  Since large spectroscopic surveys of purely
magnitude-limited samples (e.g., VVDS, Le F\`{e}vre et al. 2004)
become progressively less efficient at probing higher redshifts, a
variety of photometric criteria have been developed to efficiently
select distant galaxies.  The application of one or combination of
several of these criteria should allow us to construct samples that
are representative for the whole galaxy population at the considered
redshift.  

The Lyman-break technique (Steidel \& Hamilton 1993) was the first to
be routinely used, identifying relatively unobscured, actively
star-forming galaxies at $z \sim 3$ based on their rest-frame UV
colors.  Similar criteria were designed to probe star-forming galaxies
at $z \sim 2.3$ and $z \sim 1.7$, referred to as BX and BM galaxies
respectively (Adelberger et al. 2004).  The advent of near-infrared
(NIR) instruments on 8-10m class telescopes encouraged the study of
NIR-selected galaxies at high redshift.  The NIR flux is less affected
by dust obscuration and small amounts of recent star formation and is
therefore a better tracer of stellar mass than the optical fluxes.
The two most commonly used color criteria in the NIR to probe distant
galaxies are based on the BzK bands (Daddi et al. 2004, identifying
galaxies at $z>1.4$) and $J-K$ color (Franx et al. 2003, designed to
select red galaxies at $z>2$).  The latter class of galaxies,
so-called Distant Red Galaxies (DRGs), are characterized by the simple
color criterion $J-K>2.3$.  They are found to be massive ($M_* \sim
10^{11}\ M_{\sun}$ for $K^{tot}_{s,Vega} \lesssim 21.5$) systems (van
Dokkum et al. 2004; F\"{o}rster Schreiber et al. 2004) and range from
dusty star-forming to quiescent types (Labb\'{e} et al. 2005; Papovich
et al. 2006; Kriek et al. 2006; Wuyts et al. 2007).

In all of the surveys mentioned above, spectroscopic confirmation is
indispensable.  The high-redshift nature of a color-selected
population can only be directly verified by measuring redshifts from
their spectra.  Apart from establishing the redshift range probed, the
presence of emission and/or absorption lines provides valuable
information on the nature of the galaxies.  Moreover, having a
spectroscopic redshift reduces the number of free parameters in
Spectral Energy Distribution (SED) modeling by one and allows for a
more accurate determination of the rest-frame colors.  Finally, the
availability of spectroscopic redshifts allows us to address the
quality of photometric redshift estimates, on which many analyses of
the high-redshift galaxy population rely.

Large samples of optically selected galaxies have been
spectroscopically confirmed and their stellar populations, metallicity
and kinematics such as large-scale outflows have been studied
extensively based on the obtained optical and NIR spectra (e.g.,
Steidel et al. 1996; Shapley et al. 2003; Erb et al. 2006).  The
samples of NIR-selected distant galaxies with spectroscopic
confirmation to date are considerably smaller, the reason being
twofold.  First, their faint nature in the rest-frame UV makes optical
spectroscopic follow-up challenging.  Second, NIR spectroscopic
follow-up (e.g., Kriek et al. 2006) is time-consuming due to the lack
of NIR Multi-object spectrographs and the brightness of the night sky
at $\lambda \gtrsim 1\ \mu$m.  Especially the number of
spectroscopically confirmed DRGs to date is limited, and their
redshifts were often obtained by surveys with other selection criteria
(e.g., Reddy et al. 2006; Conselice et al. 2007).

In this paper, we report on optical spectroscopic follow-up of DRGs,
extending initial results by van Dokkum et al. (2003, hereafter vD03).
The sample is defined in \S\ref{sel.sec}.  In
\S\ref{observations.sec}, we give an overview of the observations,
followed by a description of the data reduction in
\S\ref{reduction.sec}.  Success rate and bias are discussed in
\S\ref{successbias.sec}.  \S\ref{zdistribution.sec} presents the
spectroscopic redshift distribution and \S\ref{photz.sec} discusses
the quality of photometric redshifts.  In \S\ref{lowvshighzDRG.sec} we
consider how the observed broad-band properties of DRGs at $z<2$
differ from their high-redshift counterparts.  Finally,
\S\ref{summary.sec} summarizes the paper.

Vega magnitudes are used throughout this paper.

\section {Sample selection}
\label{sel.sec}

\subsection {Pure $J-K$ selected sample}
\label{pure.sec}
During 9 observing runs from February 2002 to November 2003 we
obtained optical spectra for NIR-selected galaxies in the following
three fields: HDFS, \1054, and CDFS-GOODS.  Very deep $J_s$ and $K_s$
imaging of the 2.5'x2.5' HDFS (Labb\'{e} et al. 2003) and the 5'x5'
field around cluster \1054 (F\"{o}rster Schreiber et al. 2006) were
obtained as part of the FIRES survey (Franx et al. 2000).  The
FIREWORKS $K_s$-band selected catalog containing 10'x15' imaging of
the CDFS-GOODS (Dickinson 2001) in 17 passbands from $U_{38}$ to MIPS
24 $\mu$m is presented by Wuyts et al. (2008).

Sources for optical spectroscopy were selected with the simple color
criterion $J - K > 2.3$ (DRGs) and, with lower priority, galaxies with
$I - H > 3.0$ and $J - K < 2.3$ were placed in the masks.  The masks
were usually shared with other high-redshift candidates and bright
fillers.  Finally, 11 sources selected by their flux excess in a
narrow-band filter centered at 4190 \AA\ were placed in one of the
masks targeting the \1054 field.  In some rare cases, targets were
selected with $J - K > 2.3$ in an older catalog, and have $J - K <
2.3$ in the final catalog.  This explains why objects \#1195 and
\#1458 from vD03 are not part of the DRG sample presented in this
paper.

A total of 64 DRGs were placed in the spectroscopic masks, all of them
having $K_{s,tot} < 22.5$.  Figure\ \ref{VKvsV.fig} illustrates their
location ({\it large symbols}) in a $V_{606} - K$ versus $V_{606,tot}$
color-magnitude diagram with respect to all DRGs with $K_{s,tot} <
22.5$ ({\it small symbols}) in the three fields.  The figure
demonstrates that the DRGs selected for optical spectroscopic
follow-up span the whole 5 magnitudes in $V_{606} - K$ color occupied
by the total DRG sample.  Furthermore, they exhibit a similar range of
$V_{606,tot}$ magnitudes, with a median $V_{606,tot}$ of 26.3.
\placefigureOne

\subsection {DRGs from other surveys}
\label{others.sec}
The CDFS-GOODS field is likely the most heavily studied deep field on
the sky.  Several spectroscopic surveys have been conducted, each with
their own selection criteria, resulting in a vast database of
spectroscopic redshifts from nearby to the most distant currently
attainable.  We cross-correlated our FIREWORKS $K_s$-band selected
catalog for the CDFS field with an up-to-date list of reliable
redshifts, most of which were provided by GOODS-FORS2 (v3.0, Vanzella
et al. 2008), the K20 survey (Mignoli et al. 2005), the VVDS survey
(Le F\`evre et al. 2004), the CXO survey (Szokoly et al. 2004), the
GMASS survey (Cimatti et al. 2008), and Fadda et al. (in prep).  For
each DRG with a matching object within a (reasonably large) search
radius of $1\farcs 2$, we checked the reliability of both the redshift
identification and the cross-correlation by eye, resulting in a list of 30
additional DRGs with spectroscopic confirmation (see Table\
\ref{others.tab}).

Since different photometric criteria were applied to select these
objects (e.g., an X-ray selection for the CXO survey), the
spectroscopically confirmed DRGs in the literature are not necessarily
representative for the whole population of galaxies with $J-K>2.3$.
We therefore decide to mark them throughout the paper as having
spectroscopic redshifts, but treat them as a seperate class, i.e.,
they are not taken into account to compute the fraction of $z<2$
interlopers or to estimate the AGN fraction based on the optical
spectra.  We note that Reddy et al. (2006) present optical
spectroscopy for 19 DRGs in the GOODS-North field, but all of them
were selected by optical (BM/BX or LBG) color criteria.  Similarly,
the spectroscopically confirmed DRGs in the Extended Groth Strip
presented by Conselice et al. (2007) all stem from the R-band selected
DEEP2 survey (Davis et al. 2003).

\section {Observations}
\label{observations.sec}
A variety of optical spectrographs on 8-10m class telescopes was used
to identify redshifts of the optically very faint DRGs: the Low
Resolution Imaging Spectrograph (LRIS, Oke et al. 1995) and DEIMOS
(Faber et al. 2003) on the W.M. Keck Telescope, FORS2 (Nicklas et
al. 1997) on VLT and GMOS (Hook et al. 2003) on Gemini South.  An
overview of the spectroscopic observations is presented in Table\
\ref{observations.tab}.

Specifications for the February 2002 run, targeting the \1054 field with LRIS,
are described by vD03.  During the other LRIS runs, the 400 lines mm$^{-1}$
grism (3400 \AA\ blaze) was used on the blue arm and the 400 lines
mm$^{-1}$ grating (8500 \AA\ blaze) on the red arm.  The D680 dichroic
was used in January 2003, whereas in March and November 2003 the D560
dichroic was inserted.  The total exposure time with LRIS, spread over
2 masks in \1054 and one in CDFS, amounted to 30.5 ks.  Series of 3 or
4 exposures (typically 1800 s each), dithered in $2\arcsec$ steps
along the slit, enabled a more efficient sky subtraction.

In January 2003, we observed \1054 with DEIMOS using a 600 lines
mm$^{-1}$ grism in conjunction with the gg495 order-blocking filter.
The exposure time was 18 ks.  Two other masks, containing a handful of
$J-K > 2.3$ objects as fillers, were exposed for 36.24 ks altogether.
For the latter the grism was blazed at 7700 \AA\ and the og550 filter
was inserted.  Similar to the LRIS observations, we dithered along the
slit.

FORS2 observations with the grism GRIS\_300V, partly in combination
with filter gg375, took place in September 2002, December 2002, March
2003 and October 2003.  A total of 88.37 ks exposure time was spread
over masks in the HDFS, \1054 and the CDFS.  The same dithering
strategy as for the LRIS spectroscopy was used.
In September 2003, we observed the HDFS with the GMOS spectrograph on
Gemini South.  In order to allow for smaller slit lengths and
consequently a larger number of objects in the mask, no dithering was
applied along the slit.  Instead, a 600 lines mm$^{-1}$ grating was
blazed at 4500 \AA\ during half of the exposures and at 4530 \AA\
during the second half.  For all DRGs observed with GMOS we obtained
28.8 ks total exposures.  One red galaxy was exposed for an additional
9.6 ks as a filler in a mask with optically brighter objects.

Using the described instrument settings, we obtained spectra for a
total of 64 DRGs.  No slits containing DRGs were lost due to failures
in the reduction process or other technical problems.  Exposure times
per object varied from a minimum of 7.9 ks to a maximum of 75.34 ks.
In the course of the 9 observing runs seeing conditions were highly
variable, ranging from $0\farcs 5$ to $2\farcs 0$, with a typical
value of $1\farcs 0$.  The 1 to 1.1$\arcsec$ wide slits gave a typical
resolution of 7.5 \AA, 3.6 \AA, 10.5 \AA\ and 4.6 \AA\ (FWHM) for
LRIS, DEIMOS, FORS2, and GMOS respectively.

\section {Reduction}
\label{reduction.sec}

Multi-object spectroscopic data obtained by LRIS, DEIMOS, FORS2 and
GMOS generally undergo the same reduction steps.  For a detailed
description of the standard LRIS reduction process, we refer the
reader to van Dokkum \& Stanford (2003).  Briefly, the observations
were divided in sessions of four dithered exposures.  We used standard
IRAF tasks to subtract the bias and apply the flatfielding and fringe
correction to each of the slit exposures.  Next, cosmic rays were
cleaned and skylines subtracted.  The wavelength calibration was based
on arc lamp images, and we used the location of a bright skyline to
apply a zero-point correction.  Finally, the 4 reduced slit exposures
were aligned, averaged, and the S-distortion was removed.

The part of the slit where the target object (and possibly a second
object) is located, needs to be masked during several reduction steps.
It is of great importance that the correct part of the slit is masked.
As the NIR-selected galaxies are extremely faint in the optical, it is
impossible to measure their positions in the slit on the raw science
frames.  We determined the object position in the slit from the mask
design and verified the predicted position for bright filler objects
on the raw science frames.  The maskwidth was set to $\sim 1\farcs 9$.

In the case of the GMOS run, where no dithering was applied, the use
of 2 gratings blazed at 4500 \AA\ and 4530 \AA\ helped to distinguish
hot pixels (at fixed CCD position) from real spectral features (at
fixed wavelength).  Nevertheless, the lack of dithering resulted in a
lower quality of the spectra.  Ten out of 64 DRGs targeted by our
survey were observed only with GMOS.

\section {Results from optical spectroscopy of DRGs}
\label{15redshifts.sec}

\subsection {Redshift measurements, success rate, and bias}
\label{successbias.sec}
DRGs, while prominent in the NIR, are generally very faint at optical
wavelengths, which probe their rest-frame UV emission.  The median
$V_{606,tot}$ magnitude of all DRGs targeted by our spectroscopic
survey is 26.3, and that of the subsample for which a redshift was
successfully identified is 25.8.  Given their faint nature in the
rest-frame UV, it comes as no surprise that continua, if detected,
have a too low signal-to-noise ratio to allow for redshift
identifications based on absorption lines.  Therefore, all
spectroscopic redshifts for DRGs in our sample are based on emission
lines.  We measured their central wavelength using the IRAF SPLOT
task.  In cases where only a single emission line was detected, we use
the following arguments to favor \Lya\ over [OII]3727 as
identification: the presence of a break (lower continuum on the blue
side of the spectral feature), and absence of features at the
wavelength where one would expect H$\beta$ and [OIII]5007 if the
emission line were [OII]3727.  The typical redshift uncertainty is
$\Delta z \sim 0.001$, as estimated from the dispersion of repeated
SPLOT measurements of emission line wavelengths of galaxies whose
emission line(s) were detected in multiple independent observations.

Out of 64 galaxies satisfying the DRG criterion without further
selection bias, the optical spectroscopic follow-up resulted in 14
redshift identifications (a success rate of 22\%).  Furthermore, NIR
spectroscopy with NIRSPEC (McLean et al. 1998) on the W. M. Keck
Telescope presented by van Dokkum et al. (2004) provided a redshift
for one targeted DRG that did not show emission lines in its optical
spectrum.  The 15 redshifts for purely $J-K$ selected DRGs are listed
in Table\ \ref{ours.tab}.  Spectroscopic redshifts obtained for 49
non-DRGs during our spectroscopic campaign are listed in Table\
\ref{nonDRG.tab}.

We investigate a possible bias of the subsample of DRGs with a
successful redshift determination in Figure\ \ref{VKvsV.fig}.  The 15
spectroscopically confirmed galaxies that were selected purely on the
basis of their red ($J-K>2.3$) color are plotted with large filled
circles.  The other DRGs targeted by our survey are marked with large
empty circles.  With smaller circles, we plot all other DRGs with
$K_{s,tot} < 22.5$ in the observed fields ({\it small empty circles})
and the subsample for which a redshift was obtained by other
spectroscopic surveys ({\it small grey circles}).  The successful
targets in our spectroscopic campaign of DRGs are biased toward
brighter magnitudes in both $V_{606}$ and $K_s$ with respect to both
the whole spectroscopically observed sample and the complete sample of
DRGs in the three considered fields.  

One could expect a bias toward brighter magnitudes based on
signal-to-noise arguments.  However, the possible presence of emission
lines makes the relation between success rate and broad-band flux less
direct.  A redshift may be more easily obtained from a faint emission
line spectrum than from a brighter absorption spectrum.  We discuss
the spectral types in \S\ref{spectype.sec}.  Remarkably, Figure\
\ref{VKvsV.fig} suggests a larger dependence of the success rate on
the $K_{s,tot}$ magnitude than on the $V_{606,tot}$ magnitude, even
though the spectra were obtained in the optical.  Out of the 10 (20)
brightest targeted DRGs in $K_{s,tot}$, a redshift was successfully
derived from the optical spectra for 60\% (45\%) of them.  Considering
the brightest 10 (20) targets in $V_{606,tot}$, the success rates drop
to 50\% (25\%).  As noted before, all redshifts were based on the
presence of emission lines.  Although caution should be taken due to
small number statistics and variable seeing conditions between the
observing runs, this might hint toward an increasing prevalence of
DRGs with \Lya\ emission with brighter $K_s$-band flux.

\subsection {Spectral types}
\label{spectype.sec}

\placefigureTwo

Figure\ \ref{oneDspec.fig} presents the 1D spectra of our successful
redshift identifications.  As stated in \S\ref{successbias.sec},
continua, if detected, have a very low signal-to-noise ratio.  Since
all spectroscopic redshifts for DRGs in our sample are based on
emission lines, we should keep in mind that we are likely dealing with
a biased representation of the whole population of galaxies with
$J-K>2.3$.  At least 22\% of DRGs show line emission bright enough to
be detected in several hours exposure time on a 8-10 m telescope.  The
remaining sources may lack emission lines, or have a redshift that
places the emission lines outside the covered wavelength range (see
\S\ref{zdistribution.sec}).  We note that, for a sample of K-bright
galaxies at $z \sim 2.3$, many of which are DRGs, Kriek et al. (2006)
found that $\sim 50$\% lack emission lines in their NIR spectra
(equivalent width H$\alpha < $ 10 \AA).

Galaxies M-203 and M-508 show [OII]3727 in emission at $z \ll 2$.  All
other spectra presented in Figure\ \ref{oneDspec.fig} feature \Lya\ in
emission, possibly in combination with interstellar absorption lines
(C-5442) or confirmed by NV, SiIV, CIV and other emission lines
indicating the presence of an AGN (C-1787, C-2659).  The presence of
\Lya\ indicates that at least a quarter of the DRGs must host regions
of star formation that are not heavily obscured, complementary to an
old underlying or dusty young population that according to SED
modeling (e.g., Labb\'{e} et al. 2005; Wuyts et al. 2007; Williams et
al. 2009) is responsible for their red rest-frame optical color.
Differences between the rest-frame UV and rest-frame optical
morphologies of DRGs also indicate that these galaxies do not have
homogeneous stellar populations (Toft et al. 2005).

As illustrated by the inset 2D GMOS spectrum of H-66 in Figure\
\ref{oneDspec.fig}, a smaller feature is visible near the \Lya\
emission line of the target, offset from H-66 in the spatial direction
by $0\farcs 35$ and in the wavelength direction by 13.7 \AA.  The
relatively high dispersion of GMOS allows for an accurate measurement
of the emission line centers: 5330.8 \AA\ (H-66) and 5317.1 \AA\
(serendipitous object).  Interpreting both lines as \Lya\ at identical
cosmological distance, the shift in wavelength corresponds to a
relative velocity of $\Delta v_r = 771$ km s$^{-1}$.  At $z=3.385$ the
projected spatial offset corresponds to 2.6 kpc.

\Lya\ at 4781 \AA\ was detected in both the LRIS and 2 FORS2 spectra of
M-1061.  However, the spectrum is offset by $1\farcs 5$ from the
predicted position in the slit as calculated from the center of the
K-band flux.  An identical offset is measured between the centers of
flux on the $B$- and $K_s$-band images.  Whether the optical and NIR
light correspond to different parts of the same galaxy, or come from
physically unrelated sources, remains uncertain.  NIR spectroscopy
could confirm the redshift of the DRG unambiguously if \Ha\ is detected
at $2.5811\ \mu$m.  At $z=2.933$ the offset of $1\farcs 5$
corresponds to 11.6 kpc.  We verified that our results would not be
affected by excluding M-1061 from our spectroscopic redshift sample.

C-1787 was also observed by Norman et al. (2002).  These authors found
that at $z=3.7$, C-1787 was the most distant type-2 QSO known at the
time of discovery, showing a bright X-ray counterpart in the 1 Ms
Chandra imaging of the CDFS.  The detection of OVI, \Lya, NV, SiIV,
NIV, CIV, HeII, and CIII in our FORS2 spectrum of the source confirms
its nature.

\subsection {AGN fraction}
\label {AGNfraction.sec}
Interpreting a detection of CIV in emission as evidence for an AGN, we
find active nuclei in 13\% of the DRGs with spectroscopic redshifts.
Under the assumption that all DRGs without redshift identification
lack emission lines in their spectra, the estimated (unobscured) AGN
fraction among the observed DRGs could be as low as $\sim 3\%$.  For
comparison, 4 out of 28 (14\%) of our spectroscopically observed DRGs
in the CDFS have a X-ray detection in the 1Ms Chandra exposure on that
field (Giacconi et al. 2002).  The X-ray detected fraction among all
DRGs with $K^{tot}_{s,Vega} < 22.5$ in the CDFS amounts to 9\%.  The
estimated AGN fraction based on our optical spectroscopy is
surprisingly low compared to the AGN fraction of 20 - 30\% implied by
recent multi-wavelength studies by Reddy et al. (2005), Papovich et
al. (2006), and Daddi et al. (2007).  This might imply a prevalence of
obscured AGN among DRGs.

\subsection {Redshift distribution}
\label {zdistribution.sec}
We next discuss the distribution of spectroscopic redshifts obtained
for DRGs.  Three questions need to be addressed.  How efficient is the
DRG selection criterion to isolate galaxies at $z>2$, for which it was
designed?  What is the typical redshift of DRGs?  And to what range of
redshifts are they confined?

\placefigureThree The solid histogram in Figure\ \ref{speczhist.fig}
shows the redshift distribution of spectroscopically confirmed DRGs
from our purely $J-K$ selected sample.  The closely hatched bar at
$1.68 < z < 1.88$ marks the region in redshift space where
spectroscopic confirmation with LRIS is complicated because for
galaxies at these redshifts [OII]3727 lies redward of the covered
wavelength range while \Lya\ has not entered the blue sensitive region
of the detector yet.  Since optical spectroscopic surveys of faint
galaxies often rely on these relatively bright emission lines as sole
redshift indicators, we caution that our spectroscopically confirmed
sample may be biased against galaxies in this redshift range, even if
they have significant [OII] or \Lya\ line emission.  The corresponding
region for the FORS2 spectrograph, whose sensitivity in the blue
reaches down to $\sim 4000$ \AA, is indicated with the widely hatched
area.  Two out of 15 sources (13\%) are located below $z=2$, at
$z=1.580$ and $z=1.189$.  The median of the purely $J-K$ selected DRGs
lies at $z=2.7$ with a distribution ranging to $z=3.7$.

Considering the DRGs whose redshifts were obtained as part of other
surveys, we find that all those with a X-ray detection (Szokoly et
al. 2004) lie above $z=2$.  Cross-correlation with the K20 survey
($K_s < 20$ selected), the VLT/FORS2 survey ($z_{850}<25$ and
$i_{775}-z_{850}$ selected), and a sample of $I-K$-selected red
galaxies by Roche et al. (2006) added 9 extra low-redshift ($z
\lesssim 1.4$) interlopers.  Cross-correlation with NIR spectroscopy
of $K_s$-selected sources by Kriek et al., IRS spectroscopy of 24
$\mu$m-selected sources by Fadda et al. (in prep), and optical
spectroscopy of 4.5 $\mu$m-selected sources by the GMASS survey added
14 spectroscopically confirmed DRGs in the redshift range $1.5<z<2.5$.
Several of the Fadda et al. (in prep) sources lie in the 'no man's
land' of optical spectroscopy ({\it hatched bars in Fig.\
\ref{speczhist.fig}}).  No bias against these redshifts is present in
their sample since the redshifts were identified from MIR spectra.
Finally, applying the $J-K$ color selection to spectroscopic samples
of the MUSYC survey and VLT/VIMOS survey (Popesso et al. 2009)
resulted in 3 more spectroscopically confirmed DRGs at $z>3$.
Combining the spectroscopic redshifts from our and other surveys, we
find that the spectroscopically confirmed DRGs at $z<2$ have a median
$K_s$-band magnitude that is 0.7 magnitude brighter and a median
$V_{606}$ magnitude that is 0.7 magnitude fainter than those at $z>2$.
A similar trend is found when studying the full DRG sample (including
those that only have a photometric redshift estimate, see
\S\ref{lowvshighzDRG.sec}).  Our result is in qualitative agreement
with Conselice et al. (2007) who studied a sample of bright
NIR-selected DRGs.  Using a combination of photometric redshifts and
spectroscopic redshifts from the DEEP2 survey, the latter being
$R$-band selected and reaching to $z=1.4$, they conclude that at the
bright end ($K^{tot}_{s,Vega} < 20.5$) 64\% of the DRGs in their
sample are located at $z<2$.  Quadri et al. (2007) also found that
their (photometric) redshift distribution of DRGs shifts toward lower
redshift when imposing a brighter $K_s$-band cut.

We note that the two low-redshift interlopers from our survey are the
faintest in $K_s$ of all spectroscopically confirmed $z<2$ DRGs.  The
suggested $K_s$-band dependence of the success rate to identify
redshifts (see \S\ref{successbias.sec}) is thus not trivially related
to a redshift dependence of the success rate.

We caution that biases against galaxies without emission lines, and
galaxies in the 'no man's land' of optical spectroscopy indicated with
the hatched bars in Fig.\ \ref{speczhist.fig} might affect the
presented redshift distribution.  Therefore, we return to the
questions raised at the start of this section in
\S\ref{lowvshighzDRG.sec}, using photometric and spectroscopic
redshifts for a complete DRG sample.

\section {Photometric redshifts}
\label {photz.sec}

In order to better address the observed and intrinsic properties, and
fraction of low-redshift ($z<2$) DRGs, we complement the spectroscopic
sample presented above with photometric redshift estimates for the
remaining DRGs in the HDFS, \1054, and the CDFS.  Although photometric
redshifts have been tested for optically-bright galaxies (Wuyts et
al. 2008; Brammer et al. 2008), the number of DRGs with redshifts in
these fields has been very limited.  The reliance on photometric
redshifts has been one of the main uncertainties in, e.g., the results
on the mass assembly over cosmic time (Marchesini et al. 2009).  In
this section, we assess the reliability of published photometric
redshifts for these fields, focussing primarily on DRGs.  We first
summarize the method and templates used to estimate redshifts from
broad-band photometry.  Next, we analyse the quality of the
photometric redshifts by comparison to the available spectroscopic
redshifts.

\subsection {Method and template sets}
\label {photz_method.sec}

Using the publicly available code EAZY\footnote[1]{Code and
documentation are available at http://www.astro.yale.edu/eazy}
(Brammer et al. 2008), photometric redshifts were derived for all
$K_s$-band selected sources in the HDFS, \1054, and the CDFS.
Briefly, the program fits a non-negative linear combination of
synthetic templates to the $U$-to-8 $\mu$m broad-band photometry.
Using the $K_s$-band magnitude as a prior, and applying a template
error function that downweights the rest-frame UV and rest-frame NIR
of the templates in the fit, one obtains a redshift probability
distribution.  We adopt as best estimate the value of the redshift
marginalized over this probability distribution (see Eq. 5 in Brammer
et al. 2008), and the confidence intervals are derived from the same
distribution.

Our template set consists of 6 principal components that, in
superposition, span the entire range of galaxy colors.  The 6
templates were constructed from P\'{E}GASE models (Fioc \&
Rocca-Volmerange 1997) as described by Brammer et al. (2008), and are
identical to those used by, amongst others, Wuyts et al. (2008),
Williams et al. (2009), Marchesini et al. (2009), and Damen et al. (2009).

\subsection {Quality of photometric redshifts}
\label {photz_quality.sec}
\placefigureFour We quantify the performance of the
photometric redshift code EAZY by Brammer et al. (2008) by a direct
comparison with the available spectroscopic redshifts (see Figure\
\ref{photz_specz.fig}).  DRGs are marked in black, with large symbols
representing objects targeted by our spectroscopic survey.  Galaxies
with $J-K<2.3$ are plotted in grey.  Their spectroscopic redshifts are
compiled from the literature on the 3 fields, carefully
cross-correlating galaxies from the spectroscopic surveys to objects
in the $K_s$-band selected catalogs and conservatively limiting
ourselves to those sources with highest quality $z_{spec}$.  The
non-DRGs for which we measured redshifts during our survey (Table\
\ref{nonDRG.tab}) are also plotted in grey.

Ideally, one algorithm and set of templates provides simultaneously
accurate redshift estimates for galaxies of different types and at a
range of cosmological distances.  Here, we focus on the $z_{phot}$
quality of DRGs, but place it in context by comparing the distribution
of $\Delta z/(1+z) = \frac{(z_{phot} - z_{spec})}{(1+z_{spec})}$ for DRGs to
that of the whole population of galaxies and the subsample at $z>2$.

The results are quantified with 3 statistical measures in Table\
\ref{photozqual.tab}: the median of $\Delta z/(1+z)$ quantifies
systematic offsets, the normalized median absolute deviation
$\sigma_{NMAD}$ (equal to the rms for a gaussian distribution) is a
measure of scatter robust against outliers.  Third, we list the
fraction of catastrophic outliers ($\Delta z/(1+z) > 5 \sigma$).

We find a tight correlation between $z_{phot}$ and $z_{spec}$ for the
DRGs, without any systematic offset, and characterised by a $0.047 <
\sigma_{NMAD} < 0.056$, depending on the precise sample definition.
There are no catastrophic outliers among the DRGs.  It is reassuring that,
despite the lack of AGN templates, the template set performs equally
well for those DRGs with an X-ray detection as for the others.  This
might mean that the optical-to-NIR SEDs of these DRGs with an X-ray
detection is dominated by stellar light, and that the AGN is obscured.

Considering all 1517 galaxies with a spectroscopic redshift, 83\%
(90\%) of which lie below $z=1.5\ (2)$, we again find a small scatter
($\sigma_{NMAD} = 0.031$) and no systematic offsets.  The fraction of
catastrophic outliers is small, 3\% of all galaxies has
$\Delta z/(1+z) > 5\sigma$.  A third of these have a value of the
quality parameter $Q_z > 3$ (see Brammer et al. 2008), whereas less
than 2\% of the sources with $\Delta z/(1+z) < 5\sigma$ was flagged
with $Q_z > 3$.  The quality of photometric redshifts remains high,
even for the subsample of galaxies at $z>2$, where $\sigma_{NMAD} =
0.055$.

We conclude that a similar high quality of photometric redshifts is
reached for the spectroscopically confirmed DRGs as for the total
galaxy sample with spectroscopy.  However, as noted earlier, the
subsample of DRGs with spectroscopic confirmation is biased toward
sources with emission lines.  NIR multi-object spectrographs that will
come on-line soon will be able to establish the $z_{phot}$ accuracy
for the DRG sample as a whole in a time-efficient manner, targeting
either rest-frame optical emission lines (e.g., Kriek et al. 2007) or
Balmer/4000 \AA\ breaks in the continuum (Kriek et al. 2006).

\section {The nature of low-redshift DRGs}
\label {lowvshighzDRG.sec}

\placefigureFive Having established confidence in the $z_{phot}$
estimates for DRGs, we can now revisit the questions raised in
\S\ref{zdistribution.sec} (what is the typical redshift and redshift
range of galaxies selected as DRGs, and how efficient is the criterion
at selecting $z > 2$ galaxies), and in addition address how the
low-redshift DRGs stand out with respect to their high-redshift
counterparts.  To this purpose, we plot the $J-K$ color of all
galaxies with $K^{tot}_{s,Vega} < 22.5$ in the considered fields
versus $z_{phot}$ ({\it empty symbols}), or $z_{spec}$ ({\it filled
symbols}) when available (Figure\ \ref{JKvsz.fig}).  The median
redshift of DRGs in this combined spectroscopic and photometric sample
is $z = 2.5$, with the central 68\% of DRGs located in the
redshift interval $1.7 < z < 3.3$.  The efficiency
of the $J-K>2.3$ criterion in selecting galaxies above $z=2$ is found
to be 74\%.  The efficiency progressively increases with redder $J-K$
color.  Only 11\% (2.6\%) of the galaxies with $J-K > $ 2.6 (2.9) was
assigned a redshift below $z=2$.  Less than half of the DRGs at $z<2$
have a $J-K$ color that is consistent at the $1\sigma$ level with
being photometrically scattered into the DRG selection window, making
it unlikely that all of the low-redshift interlopers are due to
photometric uncertainties.  It is evident from Fig.\ \ref{JKvsz.fig}
that the high efficiency of the DRG criterion to select $z>2$ galaxies
does not mean that all of the $K_s$-selected galaxies at $z>2$ have
such red $J-K$ colors.  The fraction of $z>2$ galaxies that are DRGs
increases with stellar mass (Wuyts et al. 2007) and reaches 69\% at
masses above $10^{11}\ M_{\sun}$ (van Dokkum et al. 2006).

\placefigureSix
We now proceed to examine the nature of DRGs at $z<2$.  First, we
consider the observed $K_s$-band magnitude of DRGs as a function of
redshift (Figure\ \ref{Kvsz.fig}).  Apart from the spectroscopically
confirmed redshifts from our ({\it large filled circles}) and other
({\it small filled circles}) surveys, we plot the other DRGs ({\it
empty circles}) in the considered fields using their photometric
redshift estimates.  Both the spectroscopic and the photometric sample
of DRGs show a correlation between $K_s$-band magnitude and redshift.
In our sample to $K_{s,tot}<22.5$, we find a median $K_{s,tot} = 20.7$
for $z<2$ DRGs, compared to a median $K_{s,tot} = 21.4$ for $z>2$
DRGs.  Consequently, the fraction of low-redshift ($z<2$) DRGs
increases toward brighter $K_s$-band magnitudes, consistent with
Quadri et al. (2007).

\placefigureSeven In order to investigate the difference in
intrinsic properties between low- and high-redshift DRGs, we plot
their rest-frame SEDs, normalized to the rest-frame $I$-band flux, in
Figure\ \ref{restSED.fig}.  Although satisfying the same observed
color criterion ($J-K > 2.3$), the populations at low- and high
redshift show markedly different rest-frame SED shapes.  The
low-redshift DRGs show low flux levels in the UV and a positive slope
of the SED at the rest-frame $I$-band.  The high-redshift DRGs instead
show a wide range in rest-frame UV slopes and have SEDs with a
declining slope at the rest-frame $I$-band (see also F\"orster
Schreiber et al. 2004).

\placefigureEight An interpretation of the difference in rest-frame
SED shapes is provided by modeling of the optical-to-MIR SEDs using
the Bruzual \& Charlot (2003) stellar population synthesis code
following the procedure described by Wuyts et al. (2007), keeping the
redshift fixed to the $z_{phot}$, or $z_{spec}$ when available.  A
maximal visual extinction of $A_V=4$ magnitudes was allowed during the
fit, adopting a Calzetti et al. (2000) attenuation law.  Wuyts et
al. (2007) demonstrated that the inclusion of IRAC photometry
(available for all DRGs in our sample) helps to break the age-dust
degeneracy and to constrain the amount of extinction in red galaxies
at $z>2$.  The estimated extinction is plotted versus redshift in
Figure\ \ref{Avvsz.fig}.  Although DRGs at $z>2$ with several
magnitudes of extinction in the V-band do exist, a trend of $A_V$ with
redshift is significant at the 99.9\% level, both for the total sample
and the subsample with spectroscopic redshifts.  The median dust
extinction of $z<2$ DRGs is $A_V = 2.2$, compared to a median value of
$A_V = 1.0$ for the $z>2$ DRGs to the same $K_{s,tot} < 22.5$ limit.
The imposed K-band limit is at least partly responsible for the
paucity of DRGs with $A_V > 2$ at $z \gtrsim 3$ in our sample.  We
tested this by taking the best-fit templates for the most obscured
($A_V > 2$) DRGs at $z < 2$, and computing the observed $K_s$-band
magnitude and $J-K$ color when redshifting the templates out to $z =
3.0$ (3.5).  The redshifted dusty galaxies still satisfy the DRG
criterion, but at $z = 3.0$ (3.5) they drop out of our $K_s$-limited
sample in 64\% (85\%) of the cases.  The lack of low-redshift DRGs
with small amounts of extinction is real, and not influenced by such a
selection effect.

We note that more than 82\% of the DRGs at $z<2$ would also be picked
up by the $I-H>3$ selection criterion for Extremely Red Objects (EROs,
McCarthy et al. 2001).  This fraction drops to about 58\% for the DRGs
at higher redshifts.  Based on Keck spectroscopy of $I-H>3$ selected
EROs, Doherty et al. (2005) inferred a dominant old stellar population
for 75\% of the ERO sample, being responsible for their red color.
Based on our SED modeling we conclude that, with the additional
constraint of $J-K>2.3$, one preferentially selects those EROs whose
large dust content is responsible for the red slope of the SED over a
large wavelength range.

\section {Summary}
\label{summary.sec}
In this paper, we presented optical spectroscopic follow-up for a
sample of Distant Red Galaxies with $K^{tot}_{s,Vega} < 22.5$ in the
fields HDFS, \1054, and CDFS.  Redshifts were identified for a total
of 15 out of 64 of the observed DRGs.  An additional 30 DRGs, though
not necessarily representative for that population, are
spectroscopically confirmed by other surveys in the CDFS.  In
addition, we release spectroscopic redshifts of 49 non-DRGs (half of
which at $z>2$), selected from the same fields.

Using 8-10m class telescopes under varying seeing conditions, we
obtain a modest success rate of 22\% only for the DRGs, increasing
toward brighter $V_{606,tot}$ and especially $K_{s,tot}$ magnitude.
Emission line spectra are more easily identified, meaning that the
spectroscopic sample is biased toward those sources with at least some
unobscured radiating gas present.  Apart from \Lya, interstellar
absorption lines are detected in one and emission lines typical for
AGN activity in two of the high-redshift DRGs.  With only 2 objects at
$z<2$ in the purely $J-K$ selected sample, we confirm that the DRG
criterion $J-K>2.3$ is an efficient means to isolate galaxies at
$z>2$, with the redshift distribution of the purely $J-K$-selected
spectroscopically confirmed sample peaking around $z \sim 2.7$.

We use the total sample of 45 spectroscopically confirmed DRGs to
address the quality of the EAZY photometric redshift code developed by
Brammer et al. (2008).  The scatter in $\Delta z/(1+z)$ is small
($\sigma \sim 0.05$), and the comparison shows no systematic offsets.
This is true, irrespective of whether we restrict the sample to DRGs
at $z>2$ or not.  The DRGs have a similar $z_{phot}$ quality as what
is measured for all 1517 galaxies with spectroscopic confirmation in
the considered deep fields, and as the subsample of 145 galaxies at
$z>2$.

Including DRGs with photometric redshifts, we find that the median of
the DRG redshift distribution is $z=2.5$, and the efficiency of the
DRG criterion to select galaxies at $z>2$ is 74\% for our sample to
$K^{tot}_{s,Vega} < 22.5$.  DRGs at redshifts below $z=2$ are
significantly more extincted by dust than those at higher redshifts.
In observed properties, they are generally characterized by having
brighter $K_{s,tot}$ magnitudes (0.7 mag brighter in the median than
$z>2$ DRGs to the same $K_{s,tot} < 22.5$ limit), and $J-K$ colors
close to $J-K=2.3$.  SED modeling implies a median dust extinction for
$z<2$ DRGs that is as high as $A_V = 2.2$.

We conclude that the DRG criterion selects preferentially galaxies at
$z>2$, but also picks up lower redshift sources that are characterized
by different SED shapes and typically brighter $K_s$-band magnitudes.
Such biases can be avoided by selecting galaxies based on their
rest-frame properties.  Such a rest-frame selection requires very good
multi-wavelength photometry and accurate photometric redshifts.

\vspace{0.42in}
We thank Lin Yan, Dario Fadda, Mariska Kriek, Jaron Kurk, Andrea
Cimatti and the GMASS consortium for providing their spectroscopic
redshift lists, and Gabe Brammer for his support with EAZY.  Stijn
Wuyts gratefully acknowledges support from the W. M. Keck Foundation.
Based on observations carried out at the European Southern
Observatory, Paranal, Chile (Program IDs 169.A-0458, 170.A-0788,
074.A-0709, 275.A-5060, 171.A-3045).  Based on observations obtained
at the Gemini Observatory, which is operated by the Association of
Universities for Research in Astronomy, Inc., under a cooperative
agreement with NSF on behalf of the Gemini partnership.  Also based on
data obtained at the W. M. Keck Observatory, which is operated as a
scientific partnership among the California Institute of Technology,
the University of California and the National Aeronautics and Space
Administration.  The Observatory was made possible by the generous
financial support of the W. M. Keck Foundation.

\begin{references}
{\small
\reference{} Adelberger, K. L., Steidel, C. C., Shapley, A. E., Hunt, M. P., Erb, D. K., Reddy, N. A.,\& Pettini, M. 2004, ApJ, 607, 226
\reference{} Brammer, G. B., van Dokkum, P. G.,\& Coppi, P. 2008, ApJ, 686, 1503
\reference{} Bruzual, G.,\& Charlot, S. 2003, MNRAS, 344, 1000 (BC03)
\reference{} Calzetti, D., et al. 2000, ApJ, 533, 682
\reference{} Cappellari, M., et al. 2009, ApJ, 704, L34
\reference{} Cimatti, A., et al. 2008, A\&A, 482, 21
\reference{} Conselice, C. J., et al. 2007, ApJ, 660, 55
\reference{} Daddi, E., Cimatti, A., Renzini, A., Fontana, A., Mignoli, M., Pozzetti, L., Tozzi, P.,\& Zamorani, G. 2004, ApJ, 617, 746
\reference{} Daddi, E., et al. 2007, ApJ, 670, 173
\reference{} Damen, M., F\"{o}rster Schreiber, N. M., Franx, M., Labb\'{e}, I., Toft, S., van Dokkum, P. G.,\& Wuyts, S. 2009, in press (astro-ph/0908.1377)
\reference{} Davis, M., et al. 2003, SPIE, 4834, 161
\reference{} Dickinson, M.,\& the GOODS Legacy Team 2001, A\&AS, 198, 2501
\reference{} Doherty, M., Bunker, A. J., Ellis, R. S.,\& McCarthy, P. J. 2005, MNRAS, 361, 525
\reference{} Erb, D. K., Shapley, A. E., Pettini, M., Steidel, C. C., Reddy, N. A.,\& Adelberger, K. L. 2006, ApJ, 644, 813
\reference{} Faber, S. M., et al. 2003, in Iye M., Moorwood, A. F. M., eds, Proc. SPIE, Vol. 4841, Instrument Design and Performance for Optical/Infrared Ground-Based Telescopes. p. 1657
\reference{} Fioc, M.,\& Rocca-Volmerange, B. 1997, A\&A, 326, 950
\reference{} F\"{o}rster Schreiber, N. M., et al. 2004, ApJ, 616, 40
\reference{} F\"{o}rster Schreiber, N. M., et al. 2006, AJ, 131, 1891
\reference{} Franx, M., et al. 2000, The Messenger, 99, 20
\reference{} Franx, M., et al. 2003, ApJ, 587, L79
\reference{} Giacconi, R., et al. 2002, ApJS, 139, 369
\reference{} Hook, I., et al. 2003, SPIE, 4841, 1645
\reference{} Kriek, M., et al. 2006, ApJ, 645, 44
\reference{} Kriek, M., et al. 2007, ApJ, 669, 776
\reference{} Kurk, J., et al. 2009, A\&A, 504, 331
\reference{} Labb\'{e}, I., et al. 2003, AJ, 125, 1107
\reference{} Labb\'{e}, I., et al. 2005, ApJ, 624, L81
\reference{} Le F\`evre, ), O., et al. 2004, A\&A, 428, 1043
\reference{} Marchesini, D., van Dokkum, P. G., Forster Schreiber, N. M., Franx, M., Labb\'{e}, I.,\& Wuyts, S. 2009, ApJ, 701, 1765
\reference{} McCarthy, P. J., et al. 2001, ApJ, 560, 131
\reference{} McLean, I. S., et al. 1998, Proc. SPIE, 3354, 566
\reference{} Mignoli, M., et al. 2005, A\&A, 437, 883
\reference{} Nicklas, H., Seifert, W., Boehnhardt, H., Kiesewetter-Koebinger, S.\& Rupprecht, G. 1997, SPIE, 2871, 1222
\reference{} Norman, C., et al. 2002, ApJ, 571, 218
\reference{} Oke, J. B., et al. 1995, PASP, 107, 375
\reference{} Papovich, C., et al. 2006, ApJ, 640, 29
\reference{} Popesso, P., et al. 2009, A\&A, 494, 443
\reference{} Quadri, R., et al. 2007, ApJ, 654, 138
\reference{} Reddy, N. A., Erb, D. K., Steidel, C. C., Shapley, A. E., Adelberger, K. L.,\& Pettini, M. 2005, ApJ, 633, 748
\reference{} Roche, N. D., Dunlop, J., Caputi, K. I., McLure, R., Willott, C. J.,\& Crampton, D. 2006, MNRAS, 370, 74
\reference{} Shapley, A. E., Steidel, C. C., Pettini, M.,\& Adelberger, K. L. 2003, ApJ, 588, 65
\reference{} Steidel, C. C.,\& Hamilton, D. 1993, AJ, 105, 2017
\reference{} Steidel, C. C., Giavalisco, M., Pettini, M., Dickinson, M.,\& Adelberger, K. L. 1996, ApJ, 462, L17
\reference{} Szokoly, G. P., et al. 2004, ApJS, 155, 271
\reference{} Toft, S., van Dokkum, P. G., Franx, M., Thompson, R. I., Illingworth, G. D., Bouwens, R. J.,\& Kriek, M. 2005, ApJ, 624, 9
\reference{} van Dokkum, P. G., et al. 2003, ApJ, 585, 78
\reference{} van Dokkum, P. G., et al. 2003, ApJ, 587, L83
\reference{} van Dokkum, P. G., et al. 2004, ApJ, 611, 703
\reference{} van Dokkum, P. G., et al. 2006, ApJ, 638, 59
\reference{} Vanzella, E., et al. 2008, A\&A, 478, 83
\reference{} Williams, R. J., Quadri, R. F., Franx, M., van Dokkum, P. G.,\& Labb\'{e}, I. 2009, ApJ, 691, 1879
\reference{} Wuyts, S., et al. 2007, ApJ, 655, 51
\reference{} Wuyts, S., et al. 2008, ApJ, 682, 985

}
\end {references}

\placetableOne
\placetableTwo
\placetableThree
\placetableFour
\placetableFive

\end {document}